\documentclass[pdflatex,sn-mathphys-num]{sn-jnl}

\usepackage{graphicx}%
\usepackage{multirow}%
\usepackage{amsmath,amssymb,amsfonts}%
\usepackage{amsthm}%
\usepackage{mathrsfs}%
\usepackage[title]{appendix}%
\usepackage{xcolor}%
\usepackage{textcomp}%
\usepackage{manyfoot}%
\usepackage{booktabs}%
\usepackage{algorithm}%
\usepackage{algorithmicx}%
\usepackage{algpseudocode}%
\usepackage{listings}%
\usepackage{subfigure}

\theoremstyle{thmstyleone}%

\theoremstyle{thmstyletwo}%

\theoremstyle{thmstylethree}%

\begin{document}

\title[Article Title]{The Quantitative and dynamical analysis of anisotropic dispersive optical solitons }

\author*[1]{\fnm{Irfan} \sur{Mahmood}}\email{irfan.chep@pu.edu.pk}
\author[2]{\fnm{Memoona} \sur{Iqbal}}\email{}

\affil*[1]{\orgdiv{Centre of High Energy Physics}, \orgname{University of the Punjab}, \orgaddress{, \city{Lahore}, \postcode{54590}, \state{Punjab}, \country{Pakistan}}}
\affil[2]{\orgdiv{ Department of Library Science and Information Management}, \orgname{ Government Graduate College for Women Baghbanpura, Lahore }, \orgaddress{ \city{Lahore}, \postcode{}, \country{Pakistan}}}


\abstract{In this article,  we  explore the integrable aspects of generalized non-linear fluid equation possessing mixed spatial and time evolutions. This non-linear model has been acknowledged as integrable tool for the non-linear  dynamical characterization of dense material such as in optical fibre, metallic ionic fluid.  This model  widely has been  applied in physical interpretations of electromagnetic propagation through perturbed charged plasma. Here,  we  use  few integrable approaches and attain  a broad range of solitary solutions  involve  deep analytical analysis. These results are furnished with computational simulations, which enrich understanding about the accurate pictorial flow of energy. We also present a bifurcation analysis for canonical coordinates to examine the impact of parameter variations  on  non-linear behavior of the systems under consideration. Subsequently,  we  perform stability analysis to evaluate the robustness of optical solitons features  in the vicinity of equilibrium points. These results establish the reliable and accurate framework for analyzing  non linear response of materials for energy propagation  by varying  physical parameters associated with this non-linear equation model.}


\maketitle

\section{Introduction}
\begin{equation}\label{PDE1}
\mathfrak{U}_{xt}-a \mathfrak{U}\mathfrak{U}_{xx}-a\mathfrak{U}_{x}^2+\mathfrak{U}_{xxxx}=0.
\end{equation}
The emergence of non-linear differential equations  in various domains of physical sciences and engineering has got substantial attention in the context to explore their integrability.  This appearance has opened a new avenue of study in theory of non-linear science  towards  developing rigorous and integrable  methods for finding their exact solutions from application point of views. In this direction, number of exact methods have been developed without approximations among these few are, the Lax-formalise, the Darboux  transformations and Hirota bilinear method \cite{ma2012hirota}. With the addition of these some other analytical  approaches can found  in litarature which very effective to construct exact solutions. Due the wide appearance of non-linear differential equations in non-linear science as in material science, condensed matter physics, particle physics, biological sciences, physical oceanography, and geophysics etc variuos short and comprehensive methods were designed as integrable tools.  Exploring  their broad applicability in non-linear science, it is  necessary and substantial  to investigate such non-linear physical models in depth.
Numerous techniques prevail  for determining exact solutions to non-linear equations.
The extended trial equation approach is one of several formulations that have been presented for 
obtaining exact solutions for non-linear partial differential equations (NLPDEs).
 including the extended trial 
equation method\cite{akram2024exact},the sine-Gordon expansion technique,\cite{kundu2021sine}homogeneous balancing process,\cite{wang2021analytical}
Application of Bernoulli’s approach on Riccati form sub-ODE
,\cite{mirzazadeh2018optical}
which based on first integrals,\cite{feng2002first} extended $-( G'/G^2 )$ expansion
technique,\cite{mahmood2023optical} Hirota bilinear method ( s-Hirota ),\cite{ma2012hirota} an extended kudryashov method,\cite{gepreel2017exact}
method of exponentiation rational functions,\cite{aksoy2016exponential}
Inverse scattering method,\cite{cakoni2014qualitative}
extended direct algebraic method,\cite{hussain2023new}
the sub-equation method,\cite{bekir2015exact}
the modified simple equation method,\cite{mirzazadeh2014modified}
the extended Jacobi’s elliptic function method,\cite{chen2005extended}
the new extended direct algebraic method,\cite{mirhosseini2020new}
sine-cosine method,\cite{yusufouglu2006solitons}
bifurcation,\cite{ma2005bifurcation}
and many others. Within fluid dynamics, shallow water wave theory examines how waves travel across a medium where the water's depth (h) is significantly smaller than its wavelength.
A wide variety of exact solutions have been reported in the literature, including trigonometric, hyperbolic, exponential, and rational forms. Additionally, researchers have explored periodic, breather wave, rogue wave, and solitary wave solutions, with particular emphasis on soliton structures such as single soliton, multi-soliton, kink-soliton, bright soliton, dark soliton, and lump soliton solutions, etc.  \cite{raza2022complexiton} - \cite{ismael2023multiple}. The general shallow water wave equation  (GSWW) equation serves as a valuable model for examining complex non-linear phenomena observed in the real world physical system. In recent year, numerous researchers have focused their efforts on analysing the GSWW equation due to its rich mathematical structure and practical relevance. Among the various analytical techniques, the riccati equation method has gained attention for its ability to transform non-linear ordinary differential equation into solvable algebraic forms, thereby facilitating the Obtaining exact solutions.  

In non-linear wave equations, Lu and Liu \cite{lu2007bifurcation} discovered that altering the bifurcation parameters can result in qualitative changes in solution behaviour, including stable centres, saddle points, or pitchfork bifurcations.
Now look at the general equation for shallow water waves:
\begin{equation}
    u_{xxxt} + \Omega u_x u_{xt} + \epsilon u_t u_{xx} - u_{xt} - u_{xx} = 0
\end{equation}
Now the wave amplitude as a function of time and space is represented by u = u(x, t).\(\Omega\neq0\),\(\epsilon\neq0\) 
are undetermined constants. The general shallow water wave equation (GSWW) provides a broad mathematical framework for explaining how non-linear waves propagate in shallow water. The basic shallow water wave equation is concerned mostly with long, small-amplitude waves in a fluid layer
filled right through with constant depth.
Conversely, the generalized form incorporates non-linear and dispersive effects of higher order which allows it to describe more complicated physical phenomena, such as solitary wave interactions, breaking in waves, the formation of localized structures kinks, bright solitons, dark solitons and other solitons. The bifurcation analysis is carried out using the theory of the planar dynamical system\cite{hosseini2023bifurcation,li2014bifurcation}. Phase Plane analysis is an ecological analysis tool that relates to dynamical systems that consist of two dependent variables (state variables)\cite{henson2012phase} -\cite{ougrinovskaia2010ode}. We also transform the generalized, a travelling wave transformation is used to transform the shallow water wave (GSWW) equation into an ordinary differential equation (ODE), and thoroughly work on analytical solutions via several ansatzs. The homogeneous balance technique facilitates the determination of solution structures, complemented by phase plane analysis for qualitative insights, and bifurcation theory is applied in investigating the qualitative dynamics, as well as the stability of solutions. We derive various solutions, both solitons and periodic waves, among others, and demonstrate them by using $2$-D, $3$-D, contour and densities plots. Such a solution not only broadens the solution arena of the GSWW equation however, also offers functional information of the inherent physics governing non-linear wave conveyance in shallow fluids.

\section{Bifurcation  Analysis}
Mathematicians that study bifurcation theory focus on how dynamical systems behave qualitatively when their parameters change\cite{jhangeer2022bifurcation,van2017combining}.
 In our study of hydrodynamic models, it provides crucial details about how solutions (such solitary waves) can develop, fragment, or alter when important parameters reach critical values.
The general Eq. for bifurcation Analysis is given below as,\[
-c \Delta^{\prime \prime \prime}-\frac{(\Omega+\epsilon) c}{2}\left(\Delta^{\prime}\right)^{2}+(c-1) \Delta^{\prime}=0
\]
Let:
\[
B = c - 1, \quad A = \frac{(\Omega + \epsilon)}{2}c
\]
\[
- c \Delta''' - A (\Delta')^2 + B \Delta' = 0
\]
by using the transformation $$ \Delta' = \Gamma.$$
and hence substituting the value in above eq. we get
\begin{align*}
- c \Gamma'' - A \Gamma^2 + B \Gamma = 0 
\end{align*}Now, set: $$
\Gamma' = \sigma, \quad \Gamma'' = \frac{d\sigma}{dt}
$$
Substituting into the equation we get:
\[
\frac{d\sigma}{dt} + \frac{A}{c} \Gamma^2 - \frac{B}{c} \Gamma = 0
\]
\[
\frac{d\sigma}{d\Gamma} = \frac{A \Gamma^2/c - B \Gamma/c}{\sigma},
\]
Critical points occur where:
\[
\sigma = 0, \quad \frac{A}{c} \Gamma^2 - \frac{B}{c} \Gamma = 0
\]
\[
 \Gamma \left( \frac{A}{c} \Gamma - \frac{B}{c} \right) = 0
\]
\[\Gamma=0, \quad\Gamma=\frac{B}{A}.\]
Hence, equilibrium points are:$$(0, 0), \quad \left(0, \frac{B}{A} \right)$$
The vector field is:
\[
R(u,v) = 
\begin{cases}
R(0,0) \\
R\left( \frac{B}{A}, 0 \right)
\end{cases}
\]
We set:
\[
P = \sigma, \quad q = \frac{A}{c} \Gamma^2 - \frac{B}{c} \Gamma
\]
 Jacobian matrix of field of vectors\(R(u,v)\) is:
\[
R(u,v) = 
\begin{pmatrix}
\frac{dP}{d\Gamma} & \frac{dP}{d\sigma} \\
\frac{dq}{d\Gamma} & \frac{dq}{d\sigma}
\end{pmatrix}
\]
The Jacobian matrix of the system \( R(\Gamma, \sigma) \) is given by:
\[
R(\Gamma, \sigma) = 
\begin{pmatrix}
0 & 1 \\
\frac{2A\Gamma - B}{c} & 0
\end{pmatrix}
\]
The three constraints are:
    \[
    u = 0, \quad u = r_1 = \frac{3 + \sqrt{9 - 8ac}}{4a}, \quad u = r_2 = \frac{3 - \sqrt{9 - 8ac}}{4a}
    \]
 They show up based on the discriminant: \( 9 - 8ac \geq 0 \) gives 3 real points.
The nature (saddle, center)
the Jacobian is used to analyze the growth of the individual etc. As such, there is only one kind of phase portrait per group of (a, c),
demonstrates all the pertinent points of singularity \( (0, r_1, r_2) \).
\begin{figure}[ht]
\centering   
\subfigure[P1]
{\includegraphics[width=63mm,height=50mm]{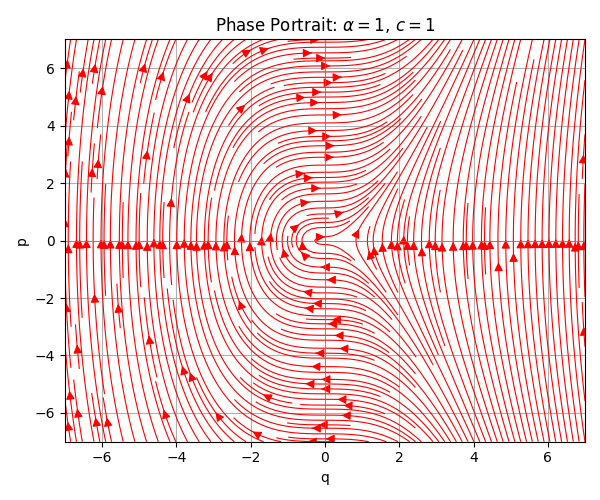}}
\hspace{0.20 cm}
 \subfigure[P2]
{\includegraphics[width=63mm,height=50mm]{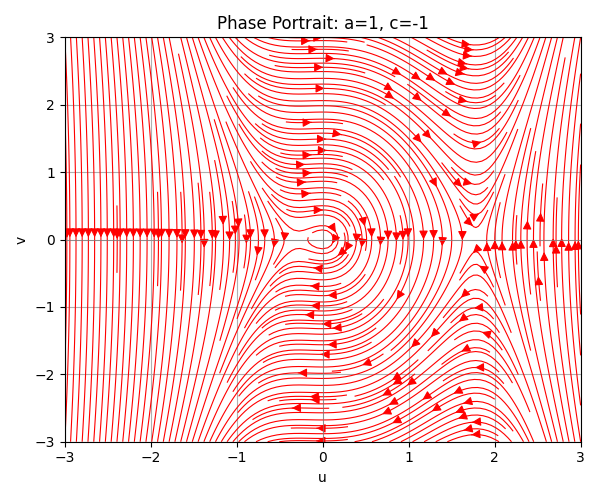}}
\hspace{0.1 cm}
\end{figure}

\begin{figure}[ht]
\centering   
\subfigure[P3]
{\includegraphics[width=63mm,height=50mm]{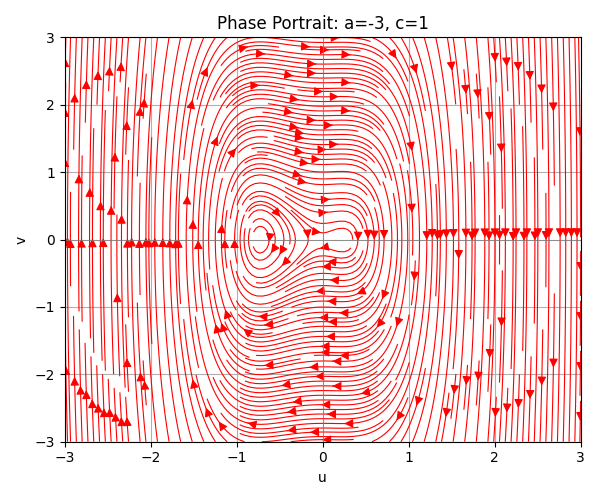}}
\hspace{0.20 cm}
 \subfigure[P4]
{\includegraphics[width=63mm,height=50mm]{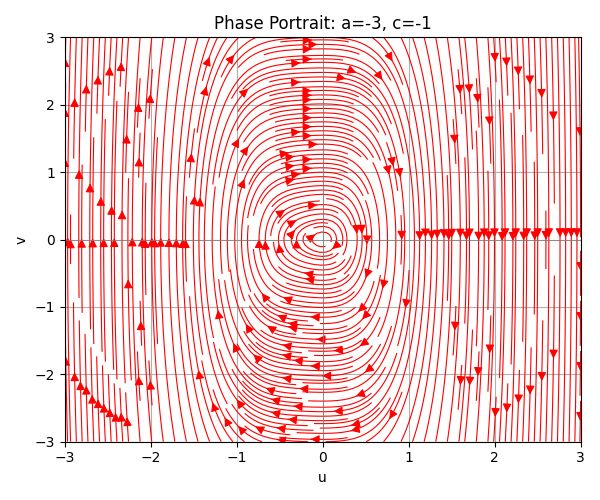}}
\hspace{0.1 cm}
\caption{The phase portrait of other values of c and a.}
\end{figure}

\section{The bunch of solitary Wave Solutions under the various methods}
\subsection{Travelling Wave Ansatz}
We apply the following transformation to reduce the equation to a travelling wave form:
\begin{equation}    
u(x, t) = \Delta(\sigma), \quad \sigma = kx + lt + d.
\end{equation}
As a result, the partial differential equation is transformed into the ordinary differential equation
 \(\sigma\). We than compute the necessary derivatives
accordingly and than put into eq(1). so we get,
\begin{align}
k^3 l\, \Delta^{\prime\prime\prime\prime} 
+ \Omega k^2 l\, \Delta' \Delta^{\prime\prime} 
+ \epsilon k^2 l\, \Delta' \Delta^{\prime\prime} 
- k l\, \Delta^{\prime\prime} 
- k^2 \Delta^{\prime\prime} = 0,
\end{align}
Now take i integration,
\begin{align}
k^3 l\, u^{\prime\prime}
+ \left( \Omega k^2 l + \epsilon k^2 l \right) \frac{u^2}{2}
- \left(k l + k^2\right) u
+ D = 0.
\end{align}
where D is an integration constant.We assume the solution of the form to solve the non-linear ODE:
\begin{align}
u = \sum_{i=0}^{n} q_i\, \Delta^i.
\end{align}
\begin{align}
\Delta' = \delta \Delta^2 + \Theta.
\end{align}
Where $q_i$ is the  constant to be determined and $\Delta$ is an auxiliary function satisfying the Riccati equation.
\subsection{Riccati Equation Solutions Using the Balance Method}
A balance between the highest order derivative and the non-linearity is achieved:
\[
u'' = u^2, \Rightarrow n = 2
\]
This implies that the solution may be a second degree polynomial, and so we shall assume that the solution may take the form:
\begin{align}
u(\sigma) = q_0 + q_1 \Delta + q_2 \Delta^2
\end{align}
where \(q_0, q_1, q_2\) are constants which must be determined, and \(\Delta\) is a function of \(\sigma\), which is obtained by ensuring it fulfils a suitable Riccati-type auxiliary equation By using eq(7-8) in eq(5).
\begin{align*}
&k^{3} l\left(q_{1}+q_{1} \Delta+q_{2} \Delta^{2}\right)^{\prime \prime}
+ \left(\frac{\Omega k^{2} l + \epsilon k^{2} l}{2}
\right)
\left( {q_{0} + q_{1} \Delta + q_{2} \Delta^{2}} \right)^{2} \\
&\quad - \left(k l + k^{2}\right)\left(q_{0}+q_{1} \Delta+q_{2} \Delta^{2}\right)
+ D = 0,
\end{align*}
$$
\begin{aligned}
& \Delta^{4}\left[6 k^{3} q_{2} \delta^{2} l+\frac{(\Omega+\epsilon)}{2} k^{2} l q_{2}^{2}\right]+ \\
& \Delta^{3}\left[2 \delta^{2} q_{1} k^{3} l+2(\Omega+\epsilon) k^{2} l q_{1} q_{2}\right]+ \\
& \Delta^{2}\left[8 q_{2} \Theta \delta k^{3} l+\frac{1}{2}\left(k^{2} l(\Omega+\epsilon)\right)\left(q_{1}^{2}+2 q_{0} q_{2}\right)-\left(k l+k^{2}\right) q_{2}\right] \\
& \Delta\left[2 q_{1}\delta \Theta+2(\Omega+\epsilon) k^{2} l q_{0} q_{1}-\left(k l+k^{2}\right) q_{1}\right]+ \\
& {\left[2 q_{2} \Theta^{2} k^{3} l+\frac{(\Omega+\epsilon)}{2} k^{2} l q_{0}^{2}-\left(k l+k^{2}\right) q_{0}+D\right]=0}. \\
& A_{1}=k^{3} l. \\
& A_{2}=\frac{1}{2}(\Omega+\epsilon) k^{2} l.\\
& A_{3}=k l+k^{2}.
\end{aligned}$$
Final Algebraic System:
\begin{align}
& 6 \delta^{2} q_{2} A_{1}+q_{2}^{2} A_{2}=0 \\
& 2 \delta^{2} q_{1} A_{1}+2 q_{1} q_{2} A_{2}=0 \\
& 8 \delta \Theta q_{2} A_{1}+A_{2}\left(q_{1}^{2}+2 q_{0} q_{2}\right)-A_{3} q_{2}=0  \\
& 2 \delta_{1} q_{1} A_{1} \Theta+2 q_{0} q_{1} A_{2}-A_{3} q_{1}=0  \\
& 2 q_{2} A_{1} \Theta^{2}+A_{2} q_{0}^{2}-A_{3} q_{0}+D=0 
\end{align}
Final answer for $q_{0}, q_{1}$, and $q_{2}$.
\begin{align}
q_{1} & =0. \\
q_{2} & =\frac{-6 \delta^{2} A_{1}}{A_{2}}. \\
q_{0} & =\frac{A_{3}-8 \delta \Theta A_{1}}{2 A_{2}}.
\end{align}
\subsection{Hyperbolic Function-based Riccati Solution}
\textbf{{Case 1: \texorpdfstring{\(\Delta(\sigma) = b_0 + b_1 \tanh(\sigma)\)}{phi(eta) = b_0 + b_1 tanh(eta)}}}
Let,
\begin{align*}
\Delta & =\sum_{i=0}^{n} b_{i} \tan h^{i} \sigma \end{align*}
We apply the homogeneous balance method to solve the Riccati equation given in Eq. (7) as follows:.
Balancing $\Delta^{\prime}$ and $\Delta^{2}$ leads to $n=1$
\begin{align}
\Delta=b_{0}+b_{1} \tan h (\sigma).
\end{align}
Substitute Eq.(17) into Eq.(7) we get,
$$
\begin{gathered}
\Delta^{\prime}=\delta \Delta^{2}+\Theta \\
\left(b_{0}+b_{1} \tanh \sigma\right)^{\prime}=\delta\left(b_{0}+b_{1} \tanh \sigma\right)^{2}+\Theta 
\end{gathered}
$$
\begin{align}
 b_{1} = \frac{-1}{\delta},\quad
b_{0} = 0.
\end{align}
By putting values of $q_{0}$, $q_{1}$,$q_{2}$ and $\Delta$ in Eq.(8), where  $
\delta \Theta = -1$.
\begin{align*}
 u(\sigma) = q_0 + q_1 \Delta + q_2 \Delta^2.
\end{align*}
\begin{align}
u=\frac{A_{3}+8 A_{1}}{2 A_{2}}-\frac{6 A_{1}}{A_{2}} {tanh}^{2} \sigma. 
\end{align}
where $\sigma=k x+l t+d$.
\begin{equation}
l = -\frac{D(\Omega + \epsilon) + k \pm \sqrt{D^{2}(\Omega + \epsilon)^{2} - 2 D k (\Omega + \epsilon) + 16 k^{6}}}{16 k^{4} - 1}.
\end{equation}
\begin{figure}[ht]
\centering   
\subfigure[3-Dim]
{\includegraphics[width=63mm,height=50mm]{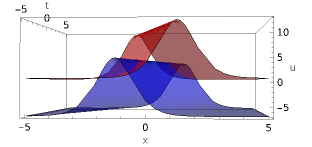}}
\hspace{0.20 cm}
 \subfigure[2-Dim]
{\includegraphics[width=63mm,height=50mm]{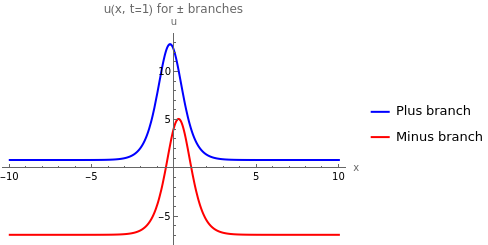}}
\hspace{0.1 cm}
\end{figure}

\begin{figure}[ht]
\centering   
\subfigure[C]
{\includegraphics[width=63mm,height=50mm]{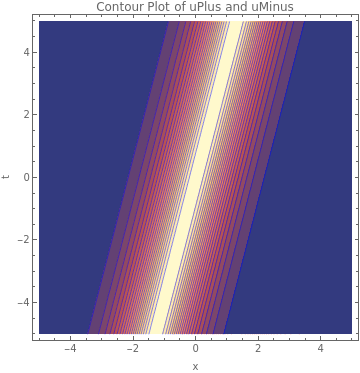}}
\hspace{0.20 cm}
 \subfigure[D]
{\includegraphics[width=63mm,height=50mm]{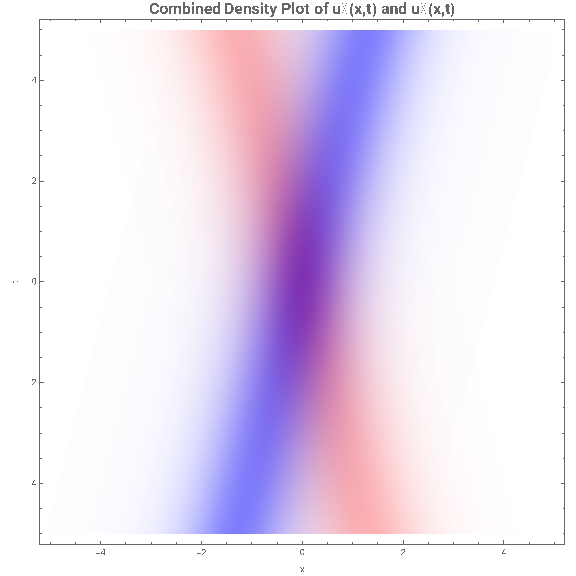}}
\hspace{0.1 cm}
\caption{(a) illustrates the three dimensional graph where k=1, d=0, $\Omega$ =0.5, $\epsilon$=0.5, D=1 and (b) Show the 2-D graph where k=1, d=0, $\Omega$=0.5, $\epsilon$=0.5, D=1 ,(c) and, (d) represent contour plot and density plot respectively }
\end{figure}
\newpage
\textbf{{Case 2: \texorpdfstring{\(\Delta(\sigma) = b_0 + b_1 \coth(\sigma)\)}{phi(eta) = b_0 + b_1 coth(eta)}}}

We choose the above form of $\Delta$ and by putting $b_{0}$ and $b_{1}$ value in above Eq. we get,
\begin{equation}
\Delta = \left( \frac{-1}{\delta} \right) \coth \sigma.
\label{eq:phi_coth}
\end{equation}
By inserting this expression into the general solution, we obtain :
\begin{align*}
& u=\frac{A_{3}+8 A_{1}}{2 A_{2}}-\left[\frac{6 A_{1} \delta^{2}}{A_{2}}\right]\left[\frac{-1}{\delta} \operatorname{coth} \sigma\right]^{2},
\end{align*}
\begin{align}
 u=\frac{A_{3}+8 A_{1}}{2 A_{2}}-\frac{6 A_{1}}{A_{2}} \operatorname{coth}^{2} \sigma.
\end{align}
\[
l = -\frac{D(\Omega + \epsilon) + k \pm \sqrt{D^{2}(\Omega + \epsilon)^{2} - 2 D k (\Omega + \epsilon) + 16 k^{6}}}{16 k^{4} - 1}.
\]
\begin{align*}   
\sigma = k x + l t + d.
\end{align*}
\begin{figure}[ht]
\centering   
\subfigure[3-Dim]
{\includegraphics[width=63mm,height=50mm]{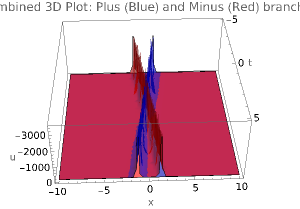}}
\hspace{0.20 cm}
 \subfigure[2-Dim]
{\includegraphics[width=63mm,height=50mm]{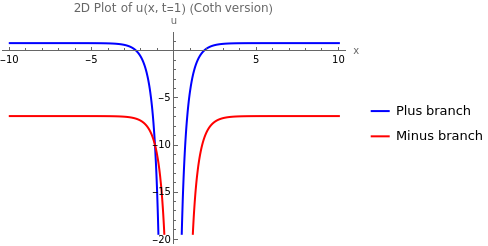}}
\hspace{0.1 cm}
\end{figure}
\begin{figure}[ht]
\centering   
\subfigure[C]
{\includegraphics[width=63mm,height=50mm]{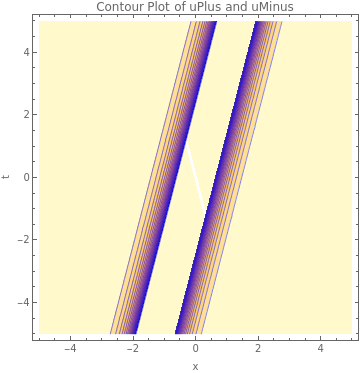}}
\hspace{0.20 cm}
 \subfigure[D]
{\includegraphics[width=63mm,height=50mm]{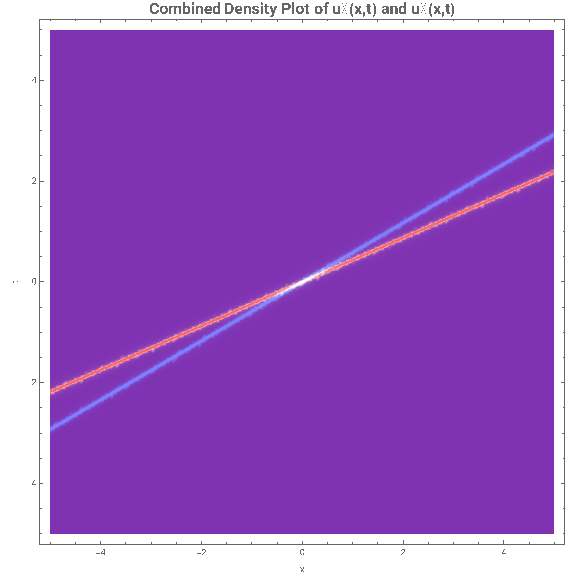}}
\hspace{0.1 cm}
\caption{(a) Depicts the 3-D graph corresponding to k=1, d=0, $\ Omega$=0.5, $\epsilon$=0.5, D=1 and (b) represents the 2-D graph where k=1, d=0, $\Omega$=0.5, $\epsilon$=0.5, D=1,(c) is a contour plot, and (d) Represents a density plot. }
\end{figure}
\newpage
\subsection{Riccati  Equation Solutions for different Parameter Regimes}
Due to the significant dependence of the system on the sign of $\Theta$ the solutions are systematically studied and presented for the following three conditions: 
We examine the three different  cases: $\Theta < 0$, $\Theta = 0$, and $\Theta > 0$
 respectively.

When $\delta=1$, the Eq.(7) has the solutions.
\begin{equation*}
\Delta =
\begin{cases}
-\sqrt{-\Theta} \tanh(\sqrt{-\Theta}\, \sigma), & \Theta < 0 \\
-\dfrac{1}{\sigma}, & \Theta = 0 \\
\sqrt{\Theta} \tan(\sqrt{\Theta}\, \sigma), & \Theta > 0
\end{cases}
\end{equation*}
\textbf{{Case I: $\Theta < 0$}}\\
when  $\Theta < 0$ then  $\Delta$ solution is;
\begin{align}
\Delta = -\sqrt{-\Theta} \tanh(\sqrt{-\Theta}\, \sigma).
\end{align}
By putting $\Delta$ in Eq.(8) we obtained,
\begin{equation}
u = \frac{A_{3} -\Theta 8A_{1}}{2A_{2}} + \frac{6A_{1}}{A_{2}} \Theta \tanh^{2}(\sqrt{-\Theta} \, \sigma).
\end{equation}
\begin{figure}[ht]
\centering   
\subfigure[3 Dim]
{\includegraphics[width=63mm,height=50mm]{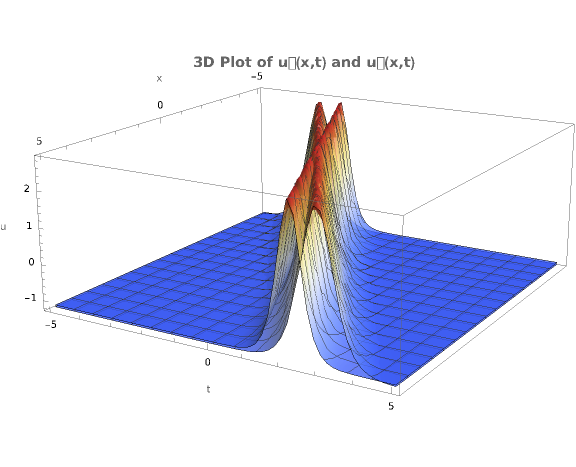}}
\hspace{0.20 cm}
 \subfigure[2 Dim]
{\includegraphics[width=63mm,height=50mm]{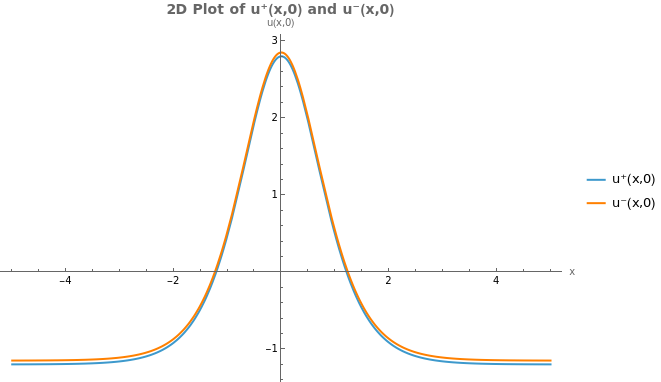}}
\hspace{0.1 cm}
\end{figure}

\begin{figure}[ht]
\centering   
\subfigure[C]
{\includegraphics[width=63mm,height=50mm]{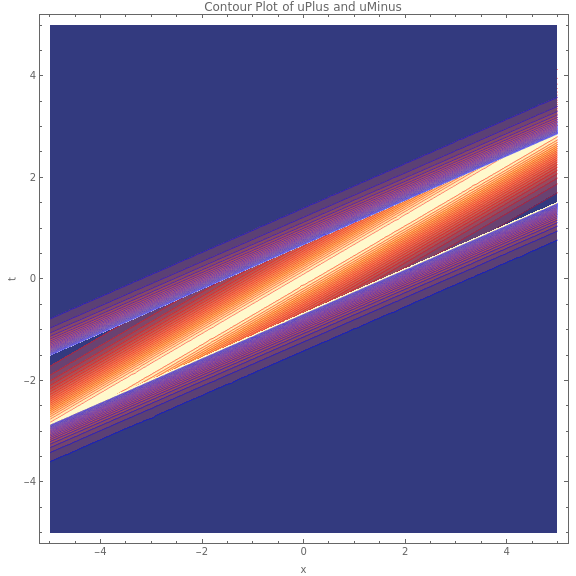}}
\hspace{0.20 cm}
 \subfigure[D]
{\includegraphics[width=63mm,height=50mm]{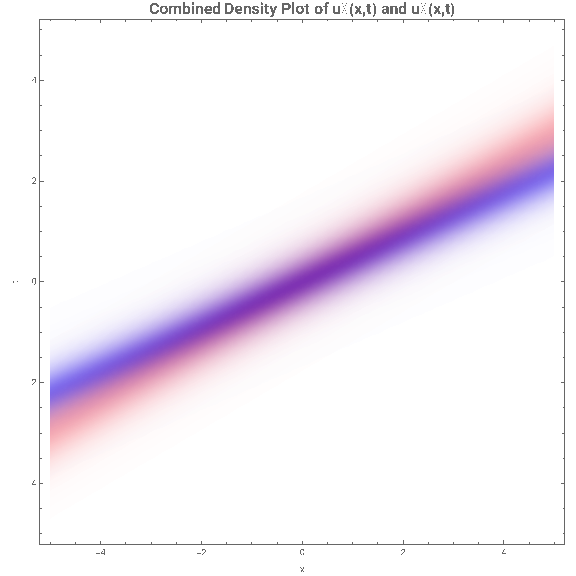}}
\hspace{0.1 cm}
\caption{(a) Three-dimensional plot is generated for k=1, d=0, $\Omega$ =1, $\epsilon$=2, $\Theta$=-1, D=1 and (b) This figure presents the 2-D graph for k=1, d=0, $\Omega$=1, $\epsilon$=2, $\Theta$=-1, D=1 ,(c) is contour plot and (d) is density plot. }
\end{figure}
\newpage
\textbf{{Case II: $\Theta=0$}}\\
Now let $\Theta=0$ then $\Delta$  value is;
\begin{align}
\Delta = -\frac{1}{\sigma}.
\end{align}
Put $\Delta$ value in Eq.(8) and we get,
$$
u=\frac{A_{3}-\Theta8 A_{1}}{2 A_{2}}-\frac{6{\delta^{2}} A_{1}}{A_{2}}\left(-\frac{1}{\sigma}\right)^{2},
$$
\begin{equation}
u = - \frac{6 A_{1}}{A_{2}} \left( \frac{1}{\sigma^{2}} \right).
\end{equation}
\[\sigma=k x+l t+d\]

\begin{figure}[ht]
\centering   
\subfigure[3-Dim]
{\includegraphics[width=63mm,height=50mm]{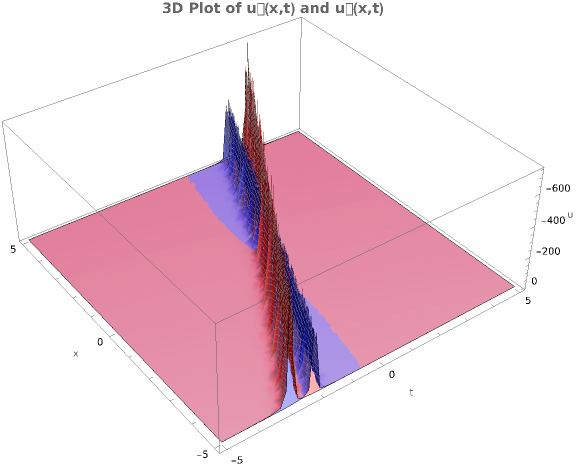}}
\hspace{0.20 cm}
 \subfigure[2-Dim]
{\includegraphics[width=63mm,height=50mm]{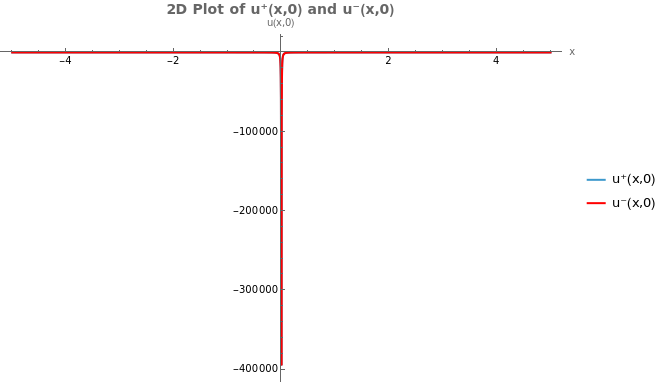}}
\hspace{0.1 cm}
\end{figure}

\begin{figure}[ht]
\centering   
\subfigure[C]
{\includegraphics[width=63mm,height=50mm]{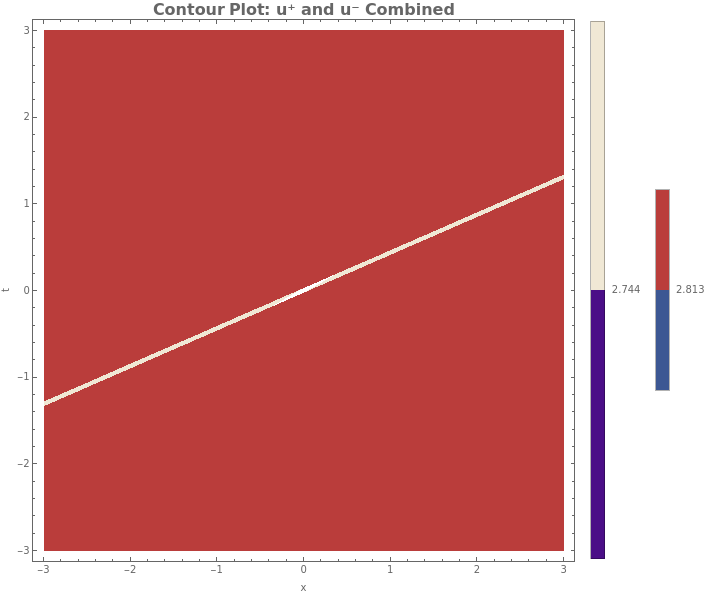}}
\hspace{0.20 cm}
 \subfigure[D]
{\includegraphics[width=63mm,height=50mm]{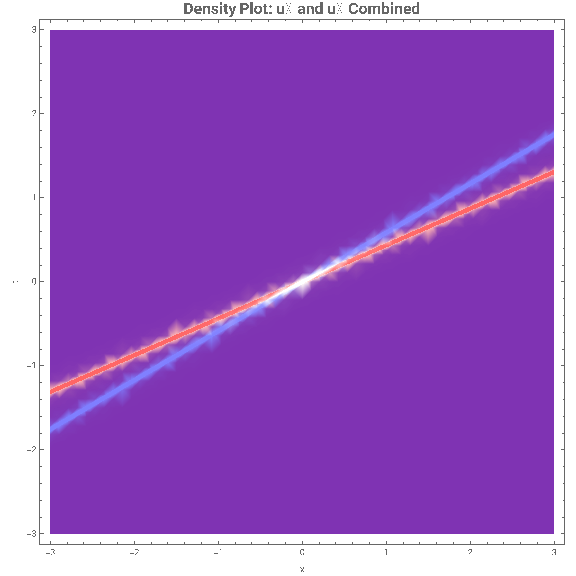}}
\hspace{0.1 cm}
\caption{(a) Depicts the 3-D graph corresponding to k=1, d=0, $\Omega$ =1, $\epsilon$=2, $\Theta$=0, D=1 and (b) represent the two dimensional graph where k=1, d=0, $\Omega$=1, $\epsilon$=2, $\Theta$=0, D=1 ,(c) is contour plot and (d) is density plot.}
\end{figure}
\textbf{Case III: $\Theta>0$}
, we have ;
\begin{align}
\Delta=\sqrt{\Theta} \tan (\sqrt{\Theta} \sigma).   
\end{align}
\begin{align}
 u=\frac{A_{3}-8 \Theta A_{1}}{2 A_{2}}-\frac{6 A_{1}}{A_{2}} \Theta \tan ^{2} \sqrt{\Theta} \sigma. 
\end{align}
\[\sigma=k x+l t+d.\]

\begin{figure}[ht]
\centering   
\subfigure[3-Dim]
{\includegraphics[width=63mm,height=50mm]{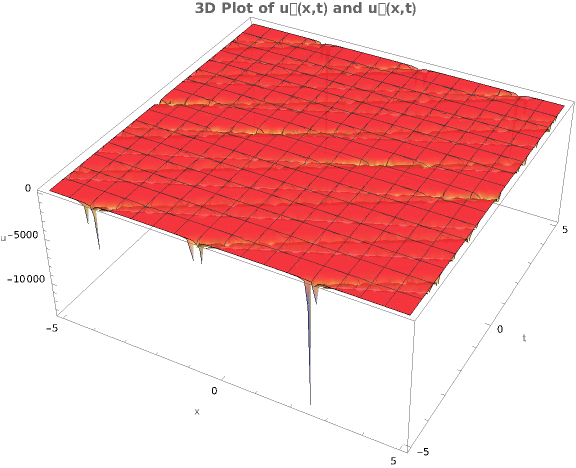}}
\hspace{0.20 cm}
 \subfigure[2-Dim]
{\includegraphics[width=63mm,height=50mm]{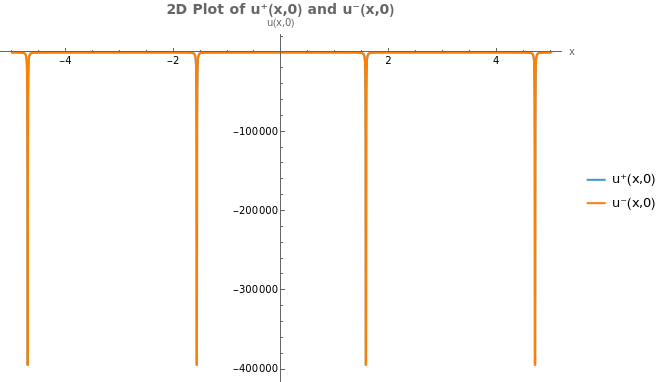}}
\hspace{0.1 cm}
\end{figure}

\begin{figure}[ht]
\centering   
\subfigure[C]
{\includegraphics[width=63mm,height=50mm]{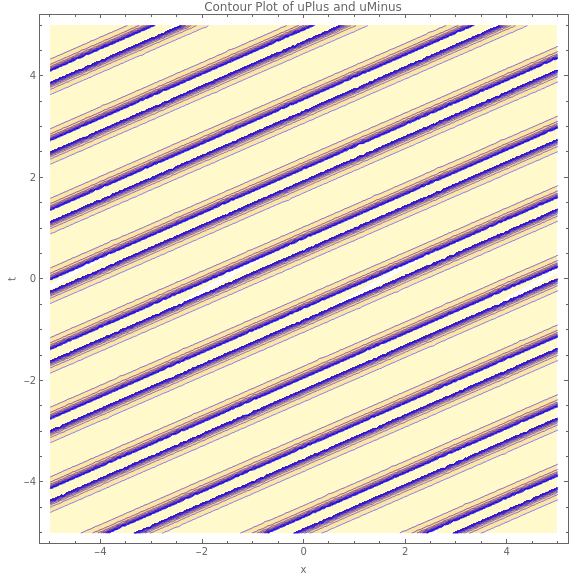}}
\hspace{0.20 cm}
 \subfigure[D]
{\includegraphics[width=63mm,height=50mm]{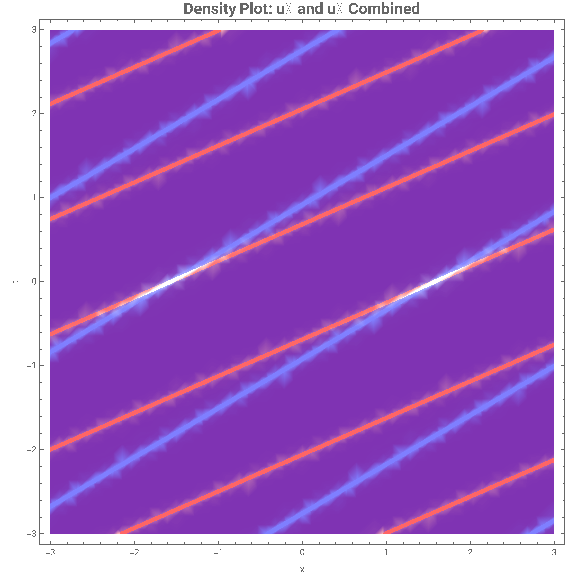}}
\hspace{0.1 cm}
\caption{(a) Illustrate the 3-D graph where k=1, d=0, $\Omega$ =1, $\epsilon$=2, $\Theta$=1, D=1 and (b) Show the two dimensional graph where k=1, d=0, $\Omega$=1, $\epsilon$=2, $\Theta$=1, D=1 , Contour and density plots are illustrated in (c) and (d), respectively}
\end{figure}

\subsection{ Approach using Auxiliary Functions}
We now want to find solutions of the Riccati Eq.(7) which we want to represent by the introduction of an ansatz in terms of auxiliary hyperbolic functions. Let,
\begin{equation*}
\Delta = \sum_{i=0}^{n} \left( A_{i} f^{i} + B_{i} f^{i} + g \right),
\end{equation*}
where,
\[
f = \frac{1}{\cosh \sigma + r}, \quad g = \frac{\sinh \sigma}{\cosh \sigma + r}.
\]
 we obtain the explicit expression of $\Delta$:
\begin{align}
\Delta = \frac{1}{2\delta} \pm \frac{\sqrt{r^{2} - 1} - \sinh \sigma}{\cosh \sigma + r}.    
\end{align}
where the parameters satisfy the constraint is \(
4\delta\Theta = -1.\)
substituting the appropriate constants, we arrive at:
\begin{equation}
u = \frac{A_{3} + 2 A_{1}}{2 A_{2}} 
- \frac{6 A_{1}}{A_{2}} 
\left(
\frac{1}{2} \pm \delta  \frac{\sqrt{r^2 - 1} - \sinh \sigma}{\cosh \sigma + r}
\right)^2.
\end{equation}
\begin{figure}[ht]
\centering   
\subfigure[3-Dim]
{\includegraphics[width=63mm,height=50mm]{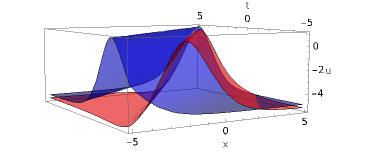}}
\hspace{0.20 cm}
 \subfigure[2-Dim]
{\includegraphics[width=63mm,height=50mm]{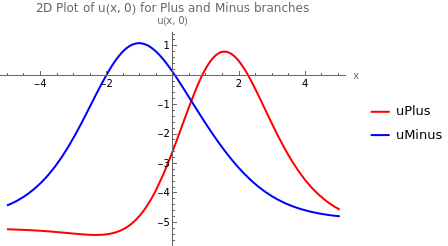}}
\hspace{0.1 cm}
\end{figure}

\begin{figure}[ht]
\centering   
\subfigure[C]
{\includegraphics[width=63mm,height=50mm]{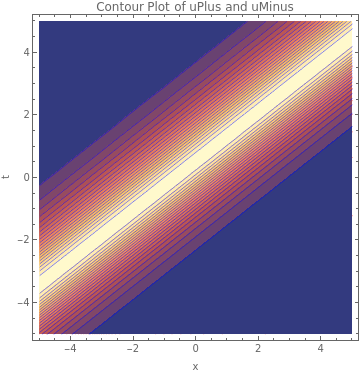}}
\hspace{0.20 cm}
 \subfigure[D]
{\includegraphics[width=63mm,height=50mm]{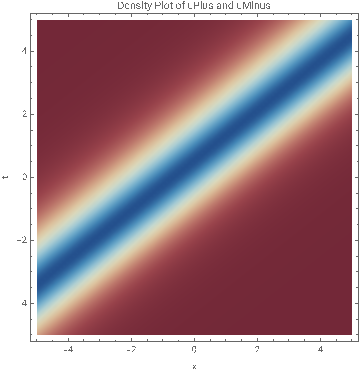}}
\hspace{0.1 cm}
\caption{(a) Show three dimensional representation where k=1, d=0, $\Omega$ =1, $\epsilon$=1, $\delta$=1, r=2, D=1 and (b) illustrate the two dimensional figures plotted for where k=1, d=0, $\Omega$=1, $\epsilon$=2, $\delta$=1, r=2, D=1 ,(c) represents the contour plot, and (d) corresponds to the density plot }
\end{figure}
\subsection{Exponential type solution Ansatz}
Let $\Delta$ be taken in  the form,
\begin{equation}
\Delta = e^{P_1 \Phi} \nu(z) + P_4(\Phi),
\end{equation}
where,
\begin{equation}
z = e^{P_2 \Phi} + P_3,
\end{equation}
and \( P_1, P_2, P_3 \) are constants to be evaluated.
Substituting Eq.(31) and (32) into the Riccati equation (7), we obtain:
\begin{align}
\Delta' &= \delta \Delta^2 + \Theta \nonumber \\
&= P_2 e^{(P_1 + P_2)\Phi} \nu' - \delta e^{2P_1 \Phi} \nu^2 + (P_1 - 2\delta P_4)e^{P_1 \Phi} \nu + P_4' - \delta P_4^2 + \Theta = 0.
\end{align}
\[
P_4 = \frac{P_1}{2\delta}, \quad \Theta = -\frac{P_1^2}{4\delta}.
\]
now Eq.(33) become ,
\begin{equation}
P_{2} \nu^{\prime}-\delta \nu^{2}=0 . 
\end{equation}
By evaluating Eq.(34), we get,
\begin{align}
\nu=-\frac{P_{1}}{\delta\left(e^{P_{1} \Phi}+P_{3}\right)} 
\end{align}
Substituting Eq.(35) and $P_{4}=\frac{P_{1}}{2 \delta}$ into Eq.(31) we have,
\begin{align}  
\Delta=-\frac{p_{1} e^{p_{1} \Phi}}{\delta\left(e^{p_{1} \Phi}+p_{3}\right)}+\frac{p_{1}}{2 \delta}
\end{align}
Following the same method as in earlier steps, we substitute the expression for  $\Delta$ value in Eq.(8), and obtain the following result:
\begin{align}
u &= \frac{A_{3} - 8 \delta \Theta A_{1}}{2 A_{2}} 
- \frac{6 A_{1} P_{1}^{2}}{A_{2}} 
\left( 
\frac{e^{P_{1} \sigma}}{e^{P_{1} \sigma} + P_{3}} + \frac{1}{2} 
\right)^2
\end{align}
Where,\(
\delta \Theta=-\frac{1}{4}\) and \(
\sigma=k x+l t+d \).
\begin{figure}[ht]
\centering   
\subfigure[3-Dim]
{\includegraphics[width=63mm,height=50mm]{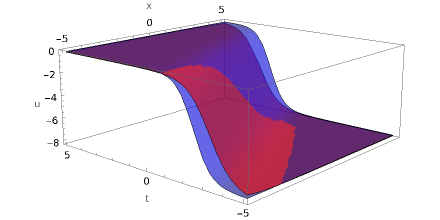}}
\hspace{0.20 cm}
 \subfigure[2-Dim]
{\includegraphics[width=63mm,height=50mm]{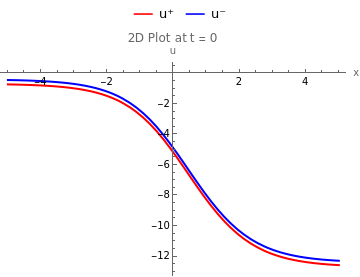}}
\hspace{0.1 cm}
\end{figure}

\begin{figure}[ht]
\centering   
\subfigure[C]
{\includegraphics[width=63mm,height=50mm]{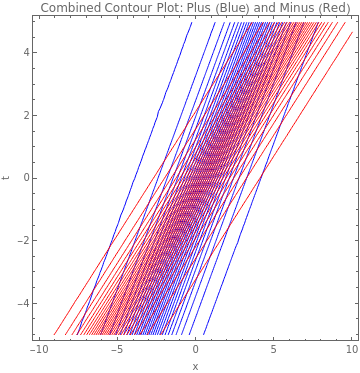}}
\hspace{0.20 cm}
 \subfigure[D]
{\includegraphics[width=63mm,height=50mm]{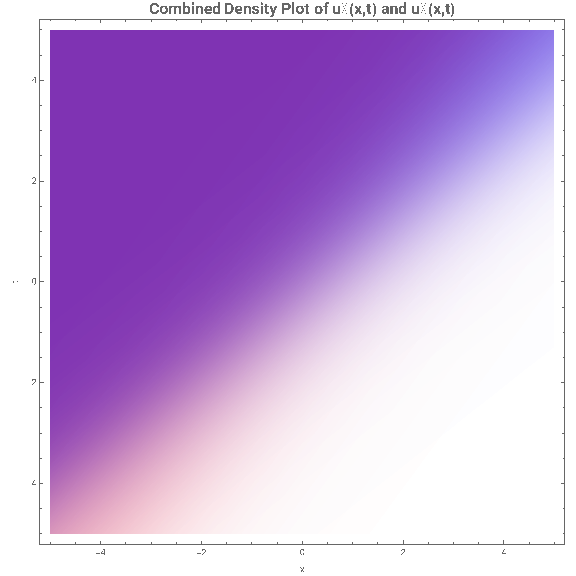}}
\hspace{0.1 cm}
\caption{(a) Illustrates the 3D graph corresponding to the parameter values: k=1, d=0, $\Omega$ =1, $\epsilon$=1, $P_1$=1, $P_3$=1, D=1 and (b) shows the 2D graph with the k=1, d=0, $\Omega$=1, $\epsilon$=2, $P_1$=1, $P_3$=1 , D=1 , (c) displays the contour plot; and (d) presents the density plot.}
\end{figure}

\subsection{ Hyperbolic Waveform Analysis and Comparative Investigation:}
\textbf{Case 1: When $P_3 = 1$} \\

Using Eq. (36), we obtain the expression for $\Delta$. By setting $P_3 = 1$, the expression simplifies to hyperbolic waveform analysis and comparative investigation.
\begin{equation}
\Delta = \frac{-P_1}{2\delta} \tanh\left( \frac{1}{2} P_1 \sigma \right)
\end{equation}
Put  $\Delta$ value in Eq.(8),
\begin{align}
u &= \frac{A_{3} - 8 \delta \Theta A_{1}}{2 A_{2}} 
- \frac{6 A_{1} P_{1}^{2}}{4 A_{2}} 
\left[ \tanh \left( \frac{1}{2} P_{1} \sigma \right) \right]^2
\end{align}

\begin{figure}[ht]
\centering   
\subfigure[3-Dim]
{\includegraphics[width=63mm,height=50mm]{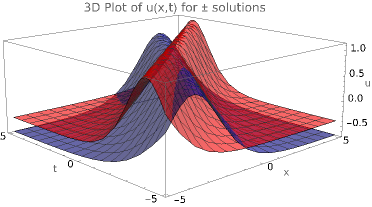}}
\hspace{0.20 cm}
 \subfigure[2-Dim]
{\includegraphics[width=63mm,height=50mm]{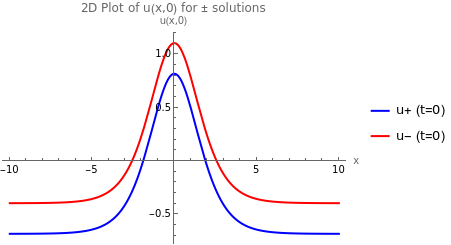}}
\hspace{0.1 cm}
\end{figure}

\begin{figure}[ht]
\centering   
\subfigure[C]
{\includegraphics[width=63mm,height=50mm]{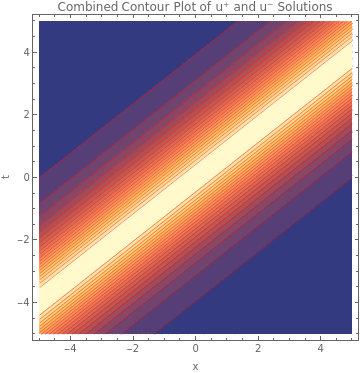}}
\hspace{0.20 cm}
 \subfigure[D]
{\includegraphics[width=63mm,height=50mm]{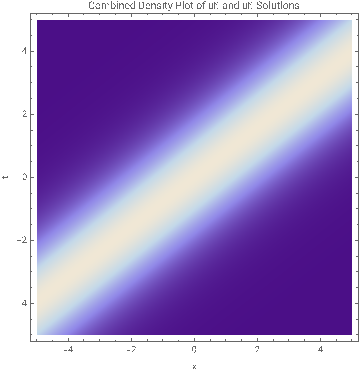}}
\hspace{0.1 cm}
\caption{(a) represent the three dimensional graph where k=1, d=0, $\Omega$ =1, $\epsilon$=1, $P_1$=1, D=1 and (b) represent the two dimensional graph where k=1, d=0, $\Omega$=1, $\epsilon$=1, $P_1$=1, D=1,(c) is contour plot and (d) is density plot.}
\end{figure}
\newpage
\textbf{Case 2:  $P_3 = -1$} \\
then
Eq.(36) is:
\begin{equation}
\Delta = \frac{-P_1}{2\delta} \coth\left( \frac{1}{2} P_1 \sigma \right)
\end{equation}
substitute this $\Delta$ value in Eq.(8);
\begin{align*}
u = \frac{A_{3} - 8 \delta \Theta A_{1}}{2 A_{2}} 
- \frac{6 A_{1} \delta^{2}}{A_{2}} 
\left( \frac{P_{1}}{2 \delta} \coth\left( \frac{1}{2} P_{1} \sigma \right) \right)^2 
\end{align*}
\begin{align}   
u=\frac{A_{3}-8 \delta \Theta A_{1}}{2 A_{2}}-\frac{6 A_{1} P_{1}^{2}}{4 A_{2}}\left[\operatorname{coth}\left(\frac{1}{2} P_{1} \sigma\right)\right]^{2}
\end{align}

\[
l = -\frac{D(\Omega + \epsilon) + k \pm \sqrt{D^{2}(\Omega + \epsilon)^{2} - 2 D k (\Omega + \epsilon) + 16 k^{6}}}{16 k^{4} - 1}
\]
where,
\[\sigma=k x+l t+d\]
\begin{figure}[ht]
\centering   
\subfigure[3-Dim]
{\includegraphics[width=63mm,height=50mm]{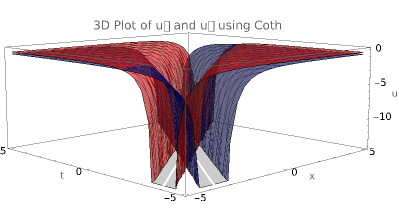}}
\hspace{0.20 cm}
 \subfigure[2-Dim]
{\includegraphics[width=63mm,height=50mm]{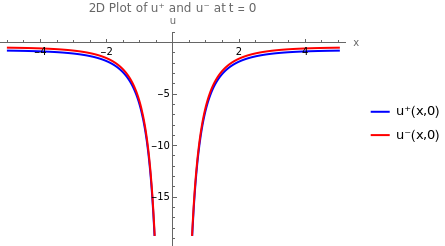}}
\hspace{0.1 cm}
\end{figure}

\begin{figure}[ht]
\centering   
\subfigure[C]
{\includegraphics[width=63mm,height=50mm]{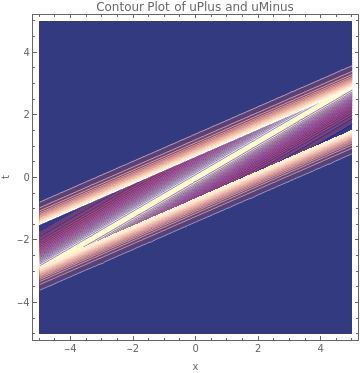}}
\hspace{0.20 cm}
 \subfigure[D]
{\includegraphics[width=63mm,height=50mm]{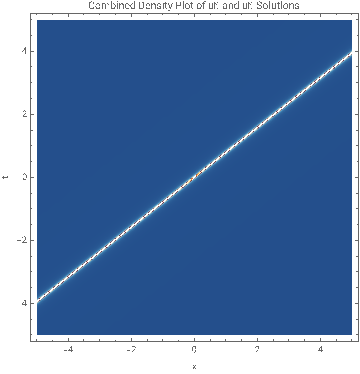}}
\hspace{0.1 cm}
\caption{    (a) A three-dimensional surface plot is presented with the parameters: $k = 1$, $d = 0$,
$\Omega = 1$, $\epsilon = 1$, $P_1 = 1$, and $D = 1$.
    (b) The corresponding two-dimensional projection is shown using the same set of parameters
    (c) A contour plot illustrating the solution profile is provided.
 (d) A density plot is shown to depict the amplitude distribute
}
\end{figure}
\section{Investigation of solution stability }
In higher-applications of shallow water (such as in Riccati-based solitons), stability analysis is used to establish the robustness of solutions (e.g. to disturbances), a condition of interest in the physical applications.
including model tsunami or coastal places.
\begin{align}
u(x,t) = P + \delta V(x,t) 
\end{align}
\begin{align*}
\delta V_{xxxt} + \Omega \delta^{2} v_{x} V_{xt} + \epsilon \delta^2 V_t V_{xx} - \delta V_{xt} - \delta V_{xx} = 0 
\end{align*}
Linearizing v:
\begin{align}
V_{xxxt} - V_{xt} - V_{xx} = 0 
\end{align}
\begin{align}
V = e^{i(kx - \Theta t)} 
\end{align}
 by substituting all values in Eq.(43) we get,
\[- k^3 \Theta e^{i(kx - \Theta t)} - (\Theta k e^{i(kx - \Theta t)}) - (k^2 e^{i(kx - \Theta t)}) = 0 
\]
\[   k^2 = \Theta (k^3 + k) 
\]
\[
\Theta = \frac{k^2}{(k^3 + k)}\] 
\begin{align}
\Theta(k) = \frac{k}{1+k^2 }   
\end{align}
This is the stability Analysis of Shallow water wave.
\section{Stability Analysis }
The stability analysis for Eq. (2) with the perturbed solution is given in this section.
 \begin{equation}
u(x,t) = l + \zeta L(x,t)
\end{equation}
where $l$ represents constant. Substituting Eq. (46) in Eq. (2), we get:==
\begin{equation}
\zeta L_{xxxt}+\Omega \zeta^2 L_x L_{xt}+\epsilon L_t L_{xx}-\zeta L_{xt}-\zeta L_{xx}=0
\end{equation}
linearize Eq. (47), we have
\begin{equation}
L_{xxxt}-L_{xt}-L_{xx}=0
\end{equation}
Suppose that the solution to Eq. (48),
\begin{equation}
L = e^{i(kx - \eta t)}
\end{equation}
where $k$ is the normalized wave number and $ \eta= \eta(k)$ is the dispersion relation. Substitute Eq. (48) for Eq. (49), compare the coefficients of the exponential function, and find $\eta$.
\begin{equation}
\eta (k)=\frac{k}{k^2+1}
\end{equation}
Since $k^2 + 1 > 0$ for all real $k$, the function $\eta(k)$ is well-defined and bounded. In particular,
\[
\lim_{k \to 0} \eta(k) = 0, \quad \lim_{k \to \pm \infty} \eta(k) = 0,
\]
and the maximum magnitude $|\eta(k)| \le 1/2$ occurs at $k = \pm 1$. \\
Moreover, the solution $L = e^{i(kx - \eta t)}$ remains purely oscillatory in time, as the exponential factor $e^{-i\eta t}$ does not produce any growth or decay. Therefore, no instability arises, and the dispersion is stable.
\section{Conclusion}
In this research, a generalized shallow water wave has been studied with the  combination of bifurcation theory, phase plane analysis, and the Riccati-based transformation methods. The non-linear system under consideration reduced to ordinary form through traveling wave transformation, thus making it possible to obtain the exact solution under different functional forms, such as hyperbolic, exponential, and auxiliary Riccati. The planar theory of dynamical systems was used to perform a complete bifurcation analysis of equilibrium points and to determine how stable they are with a Jacobian and eigenvalue analysis.Complex dynamical behaviours, including the potential for chaos, emerged in two- and three-dimensional depictions of phases with different parameters. The linearized form also confirmed the stability analysis of the solutions, which is a crucial point in the modelling of the physical phenomenon expounded in solitary wave propagation in the fluid system. A particularly notable features as  large bright and dark soliton family solutions with specific wave profiles  very accurately managed by adjusting non-linear coefficients especially the Kerr nonlinearity. The analysis presented in this paper is a strong analytical framework in learning the dynamics of nonlinear waves in hydrodynamic systems and which is transferable to other nonlinear systems. In future studies,the authors will adopt more  well design methods, integrability methods and numerical integration to study how two or more different soliton shapes interact and examine wave modulation effects across variable coefficient and higher dimensional systems.
\section*{Acknowledgement}
Our thanks to  
University of the Punjab providing facilities to complete this project.

\section*{Conflicts of Interest}
No Conflict of Interest

\section*{Authors Contribution}
All authors contributed equally. 

\section*{Funding}
No funding.
\bibliography{sn-bibliography}

\end{document}